\documentclass[fleqn,12pt,twoside]{article}
\usepackage[headings]{espcrc1}

\readRCS
$Id: espcrc1.tex,v 1.2 2004/02/24 11:22:11 spepping Exp $
\usepackage{graphicx}
\usepackage[figuresright]{rotating}

\newcommand{\AmS}{{\protect\the\textfont2
  A\kern-.1667em\lower.5ex\hbox{M}\kern-.125emS}}

\title{Jet measurements in the ALICE experiment at the LHC}

\author{J.L. Klay\address[LLNL]{Lawrence Livermore National Laboratory, \\ 
        Livermore, California 94550 USA} for the ALICE Collaboration}
       
\begin{document}

% typeset front matter
\maketitle

\begin{abstract}
Jet tomography probes provide a means to explore the properties of highly compressed and excited nuclear matter created in heavy ion collisions.  The capabilities of the ALICE experiment, with its electromagnetic calorimeter (EMCal) upgrade, to trigger on and reconstruct jets in p+p and Pb+Pb collisions at $\sqrt{s_{NN}}$=5.5 TeV are presented.  
\end{abstract}

\section{Discovery of jet quenching}
For the past six years, experiments at the Relativistic Heavy Ion Collider (RHIC) have been
investigating the properties of highly compressed nuclear matter produced in Au+Au collisions at 200 GeV.  The transition of normal nuclear matter to a state of deconfined quarks and gluons, or quark-gluon plasma (QGP) has long been predicted by Quantum Chromodynamics (QCD) at high temperatures and/or densities.  Until the advent of RHIC, the conditions necessary for such a transition are thought to only have existed at the birth of the Universe, a few microseconds after the Big Bang, and perhaps in the cores of neutron stars.  

Now there is compelling evidence, primarily in the form of jet suppression in central Au+Au collisions at $\sqrt{s_{NN}}$=200 GeV, that a dense, strongly coupled medium is formed in nuclear collisions in the laboratory.  The extraordinary jet suppression observations include a factor of five suppression of the production of high transverse momentum hadrons and the disappearance of the away side of back-to-back jets in central Au+Au collisions compared to p+p and d+Au collisions\cite{dAuPapers}.  The back-to-back jets are produced from the fragmentation of a pair of hard-scattered quarks or gluons (partons) in the initial moments of the collision.  However, as a parton propagates through the surrounding medium, it loses energy, resulting in a modified distribution of fragmentation hadrons.

Theoretical calculations of the interactions which cause partonic energy loss primarily focus on coherent medium-induced gluon radiation or gluon bremsstrahlung, with the magnitude of the energy loss being  sensitive to the color-charge density of the medium.  There are a variety of frameworks in which
these calculations have been carried out, including multiple soft scattering (BDMPS)\cite{BDMPS}, few hard scatterings (GLV)\cite{GLV}, twist expansion\cite{WangGuo} and a path integral approach\cite{Zakharov}.  While the details of the various models differ, their total energy loss is similar for comparable medium properties\cite{SalgadoWiedemann}.

Recent studies by the STAR experiment extend two-particle azimuthal correlations to lower associated hadron momenta ($p_T>$ 0.15GeV/c) and to higher $p_T$ trigger (8$<p_{T}<$15 GeV/c) and associated hadrons.  For the high $p_T$ studies, although still suppressed relative to the near-side correlations, the recoiling away-side jets re-emerge from the medium.  Quantifying the evolution of this re-emergence may help constrain estimates of the energy density.  The lower $p_T$ studies show the first hints of what happens to the energy which is deposited in the medium by the propagating parton.  The away-side correlation is enhanced in central Au+Au collisions relative to p+p and d+Au.  In fact, the strength and shape of the correlation may indicate that the recoiling jet thermally equilibrates with the surrounding medium or induces shock waves in the matter\cite{MachCone}.  

While the available data are sufficient to make qualitative estimates of the properties of the medium, the fragmentation and geometrical biases associated with these measurements limit their ability to distinguish among the different mechanisms for partonic energy loss.  Truly quantitative assessments of these features require more complete measurements of the jet energy distributions at larger parton $p_T$, where pQCD calculations and theoretical tools are more reliable. 

\section{The LHC: Jet factory}

\begin{figure}[htbp]
\includegraphics[width=0.5\textwidth]{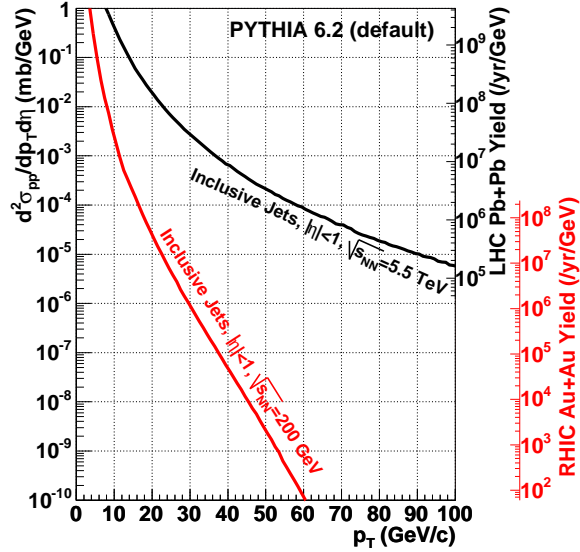}
\caption{\label{Fig:lhcjets} Comparison of jet production cross-sections in p+p (left axis) and annual yields (right axes) in Au+Au (Pb+Pb) at RHIC (LHC) using {\sc pythia}\cite{Pythia}.  The yields are for minimum bias Pb+Pb (Au+Au) collisions assuming 10$^6$ s (10$^{7}$ s) running time and 0.5 mb$^{-1}$ s$^{-1}$ (5.0 mb$^{-1}$ s$^{-1}$) luminosity at LHC (RHIC).}
\end{figure}

When the CERN Large Hadron Collider (LHC) starts colliding Pb ion beams in 2008-09, jet tomography studies of nuclear matter in extremis will enter a new era.  The jet production cross-sections at the LHC, where the collision energy is 30$\times$ higher than RHIC, are expected to be many orders of magnitude larger (Fig.~\ref{Fig:lhcjets}), while the underlying event multiplicities are expected to grow only logarithmically.  Large statistical samples of jets at RHIC energies ($E_T\sim$20-40 GeV) as well as those at 100-200 GeV will provide the dynamic range necessary to explore the mechanisms underlying partonic energy loss in detail.  With such a broad range of parton energies, it is expected that at LHC, in contrast to RHIC, high $E_T$ jets may be identified unambiguously above the background of uncorrelated hadrons from the underlying event on an event-by-event basis.  Such measurements represent a considerable improvement over the indirect jet measurements of single hadron spectra and two-particle azimuthal correlations accessible at RHIC.  However, comparisons of the same measurements at RHIC and LHC in overlapping kinematic ranges will also be important for establishing a deeper understanding of the medium created at both energy regimes.

Probably the most important observable that will become accessible at LHC is the measurement of the coincidence of a high $E_T$ jet recoiling against a single photon.  Since the photon does not interact with the medium, its energy directly reflects the recoiling parton's energy.  Therefore the fragmentation of such jets can be precisely explored and details about the jet modification can be measured and compared to theoretical expectations.  The $\gamma$-jet coincidence rate in heavy ion collisions at LHC is expected to be sufficiently large that robust, statistically significant measurements out to substantial jet transverse momenta will be possible\cite{CERNYellowReportJets}.  

In addition to the strong enhancement in the jet production cross-section, heavy quark production is expected to be substantially enhanced relative to RHIC.  Per central Pb+Pb collisions at 5.5 TeV, $\sim$250 ($\sim$7) $c\overline{c}$ ($b\overline{b}$) pairs are expected, compared with $\sim$10 ($\sim$0.05) per Au+Au collision at 200 GeV\cite{YellowReportHeavyQuarks}.   The fragmentation of heavy quarks produces jets containing heavy mesons which have a large semi-leptonic decay branching ratio to single electrons.  Therefore, the presence of a high transverse momentum electron within a jet can be used to tag the jet as coming from a heavy quark parent parton\cite{CDFbjets}.   This is an attractive avenue for exploring the flavor and mass dependence of partonic energy loss by comparing heavy quark jets to inclusive jets (dominated by gluon fragmentation) in Pb+Pb and p+p collisions\cite{KlaySQM04}.

The LHC will bring an abundance of jet quenching observables.  The broad range of theoretical predictions of significant medium-induced energy loss\cite{MLLA,SalgadoWiedemannJetShape} and the rich phenomenology of jet measurements already observed at RHIC underscore the importance of measuring the full momentum spectrum of jet fragments from $p_{T}\sim$ 100 MeV to 100 GeV.  This is the only way to study both the mechanisms of partonic energy loss and the response of the medium to the deposited energy.  Thus, a detector with good triggering, robust tracking, detailed particle identification and reasonable jet energy resolution are essential to fully exploit the richness of hard probes at the LHC.
 
\section{``A Large Ion Collider Experiment'' (ALICE)}

\begin{figure}[htbp]
\includegraphics[width=0.75\textwidth]{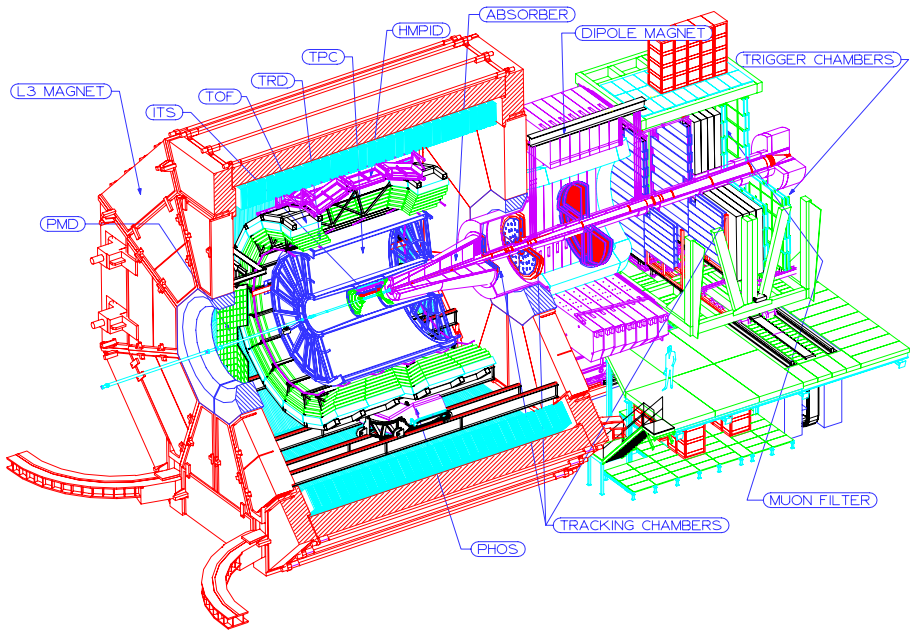}
\caption{\label{Fig:alice} Layout of the ALICE detector.}
\end{figure}

ALICE is the only LHC experiment dedicated to the study of heavy ion collisions.  Its design has been optimized for very high multiplicities to allow for precision measurements of the collision products over a broad range of transverse momenta and rapidity\cite{ALICEPPR}.  Fig.~\ref{Fig:alice} shows the layout of the ALICE detector, designed to measure and identify mid-rapidity hadrons, leptons and photons from $\sim$100 MeV to $\sim$100 GeV, along with a dedicated muon spectrometer at large rapidities (-4.0$< \eta <$-2.4).  

The central rapidity region ($|\eta|<$0.9) is composed of a complex system of sub-detectors within a 0.5 T solenoidal magnetic field.  The tracking of charged hadrons and leptons is accomplished with a set of high-granularity detectors: an Inner Tracking System (ITS) consisting of six layers of silicon detectors, a large-volume Time-Projection Chamber (TPC) and a high-granularity Transition-Radiation Detector (TRD).  In addition to the particle identification provided by the tracking detectors, there is also a high resolution array Time Of Flight (TOF) detector and a Cherenkov-based High-Momentum Particle Identification Detector (HMPID).  Photons are measured in a small-acceptance crystal PHOton Spectrometer (PHOS) calorimeter.

While the design of these subsystems was optimized for the extreme case of $dN/dy\sim$ 8000 in central Pb+Pb events, ALICE is capable of and will measure collisions of lighter ions and proton beams over a range of collision energies.  The tracking efficiency in the central barrel is expected to be over 90\%.  Despite the modest magnetic field, the fine position resolution and large lever arm of the tracking system give ALICE a momentum resolution of $\Delta p/p\sim$6\% at 100 GeV.  The addition of a large-area electromagnetic calorimeter for neutral energy and jet measurements would complete the mid-rapidity capabilities of ALICE to measure jet quenching at the LHC. 

\subsection{ALICE Electromagnetic Calorimeter (EMCal)}

The ALICE ElectroMagnetic Calorimeter (EMCal) is an upgrade proposed by a consortium of U.S. and European institutions intended to provide ALICE with high $p_T$ photon, electron and jet triggering, as well as more complete jet energy measurements and extended electron and photon identification capabilities\cite{EMCALTP}.  The most important features of the EMCal are the measurement of the neutral portion of jet energy over wide acceptance, and an efficient and unbiased fast trigger (Level 0/1) for high energy jets. It is a large Pb-scintillator sampling calorimeter with cylindrical geometry, located adjacent to the ALICE magnet coil at a radius of 4.5 meters from the beamline, outside the central tracking detectors. Its coverage in phase space is $|\eta|<$0.7 and $\Delta\phi$=110 deg, positioned opposite in azimuth to the PHOS detector. The EMCal is segmented into 12,672 projective towers, each covering $\Delta\eta \times \Delta\phi \sim$ 0.014 $\times$ 0.014. Readout fibers are configured in a Shashlik geometry and are coupled to an Avalanche Photodiode (APD) sensor. The EMCal provides fast triggers (level 0 and 1) for photons, electrons, and jets.  

\section{Jet performance studies with ALICE+EMCal}

The EMCal design has been evaluated for the four main aspects of its physics program: jet trigger, jet reconstruction, direct photon measurements and high $p_T$ electron identification in p+p and Pb+Pb collisions.  The results are documented in a Technical Proposal\cite{EMCALTP}.  Here the performance for jet triggering and reconstruction are summarized.  All studies were conducted with simulated p+p and Pb+Pb collisions using the {\sc pythia}\cite{Pythia} and {\sc hijing}\cite{Hijing} Monte-Carlo event generators.  

The EMCal can provide fast triggering capabilities for high $p_T$ $\pi^{0}$ and jets via energy sums over various patch sizes.   Based on detailed studies, the EMCal is found to achieve trigger enhancements of a factor 10-50, depending on collision system and luminosity, which significantly increase the ALICE kinematic and statistical reach, out to 200 GeV ($>$100 jets) for each one month of Pb+Pb collisions expected per year at the LHC.  The increased reach is crucial for mapping the energy evolution of jet quenching, a central goal of the ALICE physics program.

In addition to triggering on jets, the information on neutral energy measured by the EMCal 
significantly improves the jet energy resolution and minimizes the jet reconstruction bias from using the tracking system alone.  Studies of the jet reconstruction performance of ALICE tracking plus EMCal have been performed using a UA1-type cone algorithm\cite{UA1Jets}.  The algorithm has been extended to estimate and subtract the large uncorrelated background of energy from the underlying event on an event-wise basis\cite{JPhysGJets}.  

Jets are formed by incorporating electromagnetic energy measured by the EMCal and hadronic energy carried by charged particles measured in the tracking chambers (with hadronic energy deposition in the EMCal corrected on a track-wise basis), into cones of radius R=$\sqrt{\Delta\phi^2 + \Delta\eta^2}$ around a high $p_T$ seed particle.  The energy carried by unmeasured particles (primarily neutrons and $K^{0}_{L}$) is estimated to be less than 10\% and is corrected for on average.   The use of the ALICE tracking detectors to determine the hadronic component of the jet energy limits the achievable energy resolution compared to a pure calorimeter jet-finder, but provides the capability to optimize the signal to background ratio in the measurement of the jet energy better than calorimeters alone.

In fact, background energy and its fluctuations represent a significant challenge to jet-finding and jet reconstruction in heavy ion collisions.  While jets with $E_{T}\sim$ 50-100 GeV in p+p collisions distribute 80\% of their charged track energy within a cone of radius $\sim$0.2\cite{CDFChargedJets}, the background energy from a Pb+Pb collision at 5.5 TeV within an area of this size may be as large as 75 GeV.   Fluctuations in the background due to impact parameter variations, statistical fluctuations due to the finite number of tracks, and dynamical fluctuations due to lower $E_T$ jets complicate matters.  Event-wise estimates of the background within an area comparable to the jet cone can reduce the influence of the impact parameter fluctuations on the background subtraction.

\begin{figure}[htbp]
\begin{minipage}[t]{77mm}
\includegraphics[width=\textwidth]{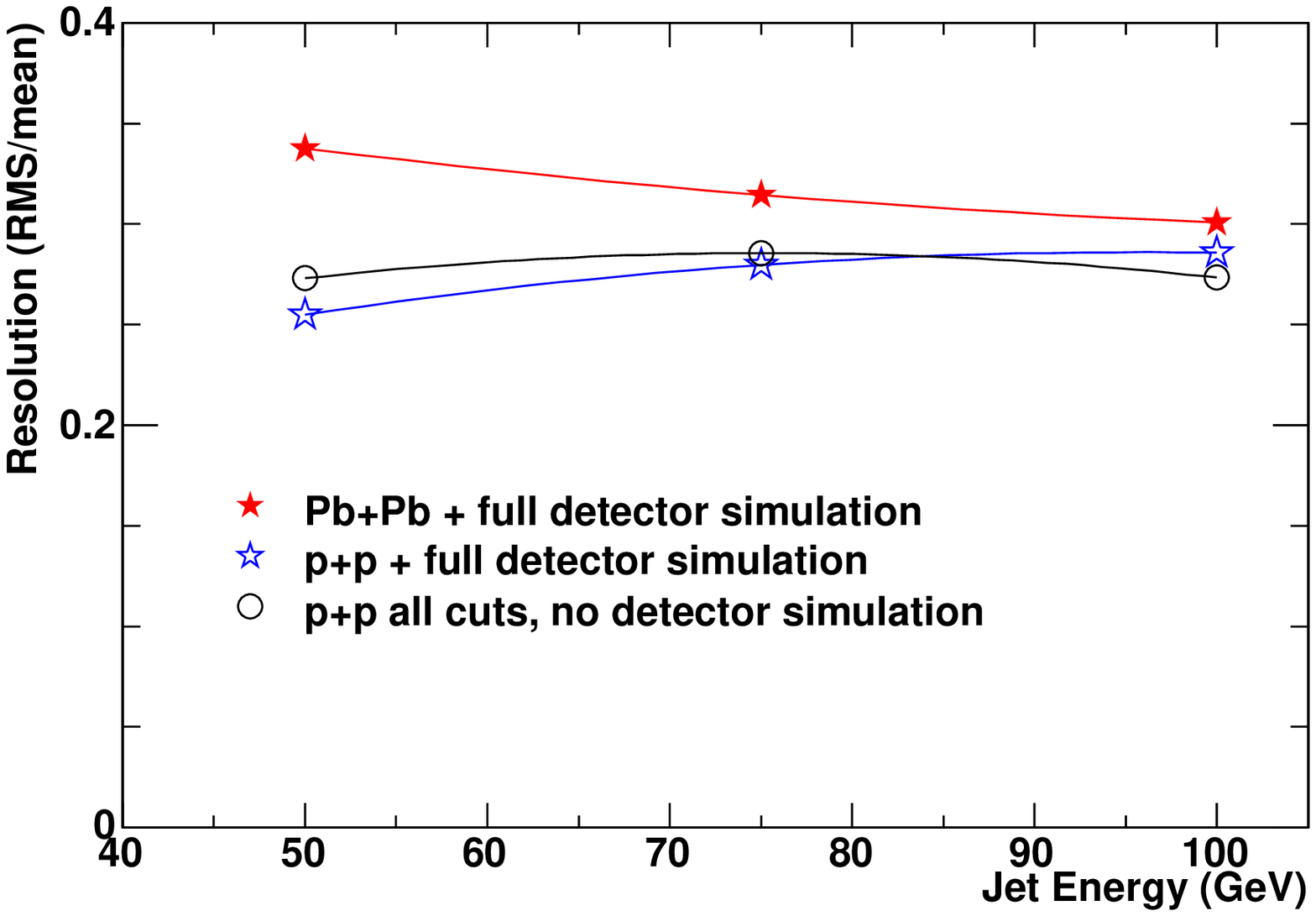}
\caption{\label{Fig:JetRes} Jet energy resolution for p+p and Pb+Pb with $p_T>$2 GeV/c and  R=0.3.  Remaining Pb+Pb background energy is subtracted on an event-wise basis\cite{JPhysGJets}.}
\end{minipage}
\hspace{\fill}
\begin{minipage}[t]{77mm}
\includegraphics[width=\textwidth]{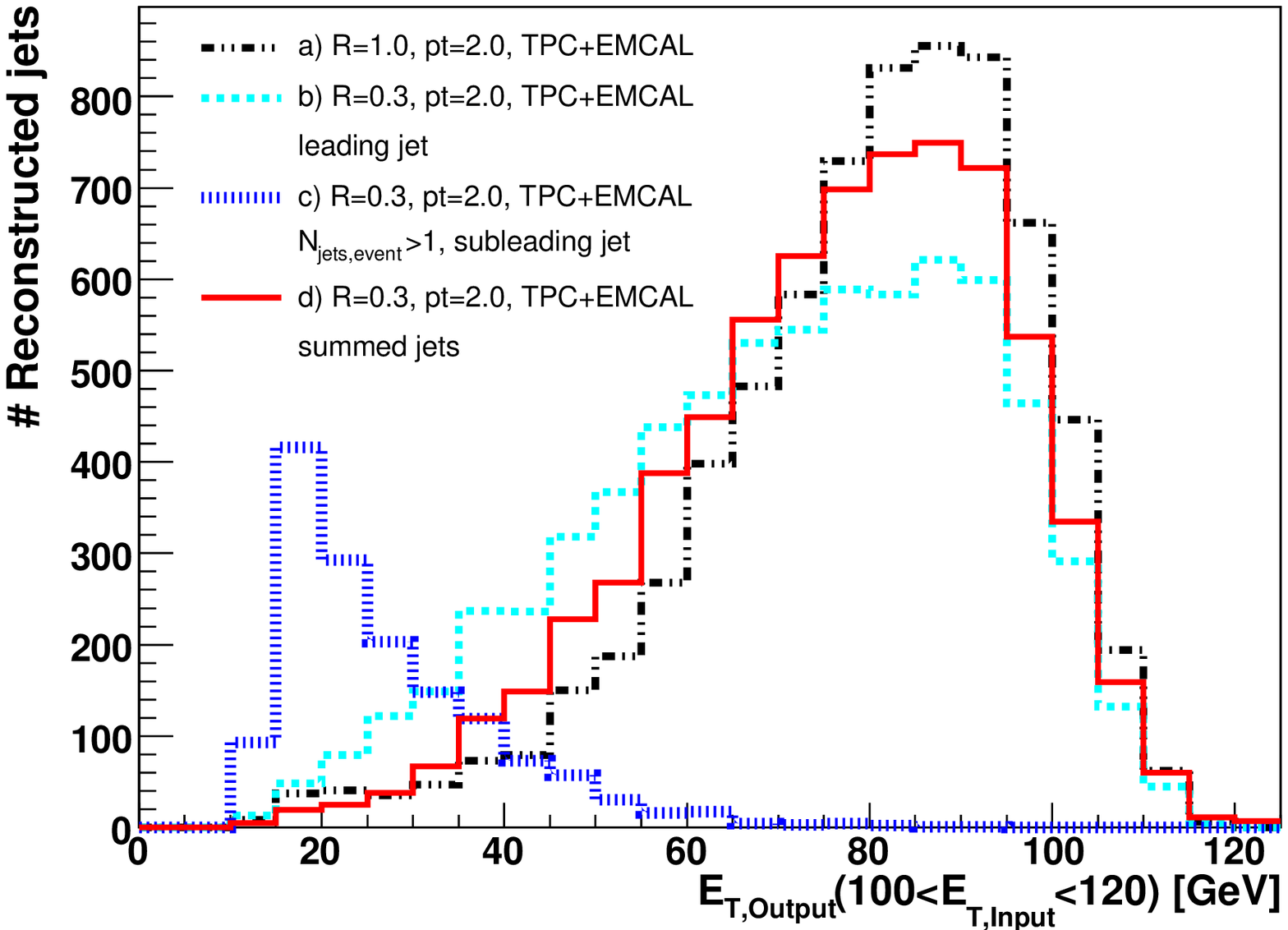}
\caption{\label{Fig:SplitJets} Reconstructed jet energy distributions for 100-120 GeV jets that illustrates the problem of jet splitting for algorithms utilizing small cone radii.}
\end{minipage}
\end{figure}

For jet reconstruction in heavy ion events, two cuts are found to be important to reduce the overall background contribution while maintaining significant jet signal: a limit on the jet cone radius R and a lower bound on the $p_T$ of tracks considered by the algorithm.  These cuts are somewhat correlated, due to the well-known angle ordering in jet fragmentation. A study with the {\sc hijing} heavy ion event generator model\cite{Hijing} shows that accepting only tracks with $p_{T}>$2 GeV/c excludes 98\% of the background tracks.  

The jet energy resolution in p+p and Pb+Pb collisions for mono-energetic jets using tracks with $p_T>$2 GeV/c and cone radius R=0.3 is shown in Fig.~\ref{Fig:JetRes} from \cite{JPhysGJets}.  The resolution in both systems is found to be comparable at large $E_T$, at approximately $\sim30\%$.  Although the strict background reduction cuts are not necessary for p+p collisions, they must be applied so that one-to-one comparisons of measured jet fragmentation functions in p+p and Pb+Pb can be made to extract the medium modifications of interest.

The results of jet-finding presented so far are for the highest energy jet found in {\sc hijing} events where one input jet of a specific energy has been embedded.  The use of the small jet cone, necessary to reduce backgrounds, mis-reconstructs jets which have a hard radiation during fragmentation because the energy is split into sub-clusters that appear as multiple jets.  Fig.~\ref{Fig:SplitJets} shows the reconstruction of 100-120 GeV {\sc pythia} jets using small cone radius and $p_T>$2 GeV/c, but without the heavy ion background.   For reference, curve (a) shows the reconstructed energy distribution for R=1, while curves (b) and (c) show the energy distribution of the highest and second highest energy jets found in the event with cone radii of R=0.3.  Curve (d) is the distribution of summed energies for the highest and second highest energy jets, which is seen to better match the distribution from the cone algorithm with large radius.  This study is a first step toward a more advanced jet algorithm which can utilize small jet cones (to reduce background) while counteracting the effects of jet splitting by re-summing nearby ``proto-jets'' in a manner similar to $k_T$ clustering algorithms.

\begin{figure}[htb]
\begin{minipage}[t]{77mm}
\includegraphics[width=\textwidth]{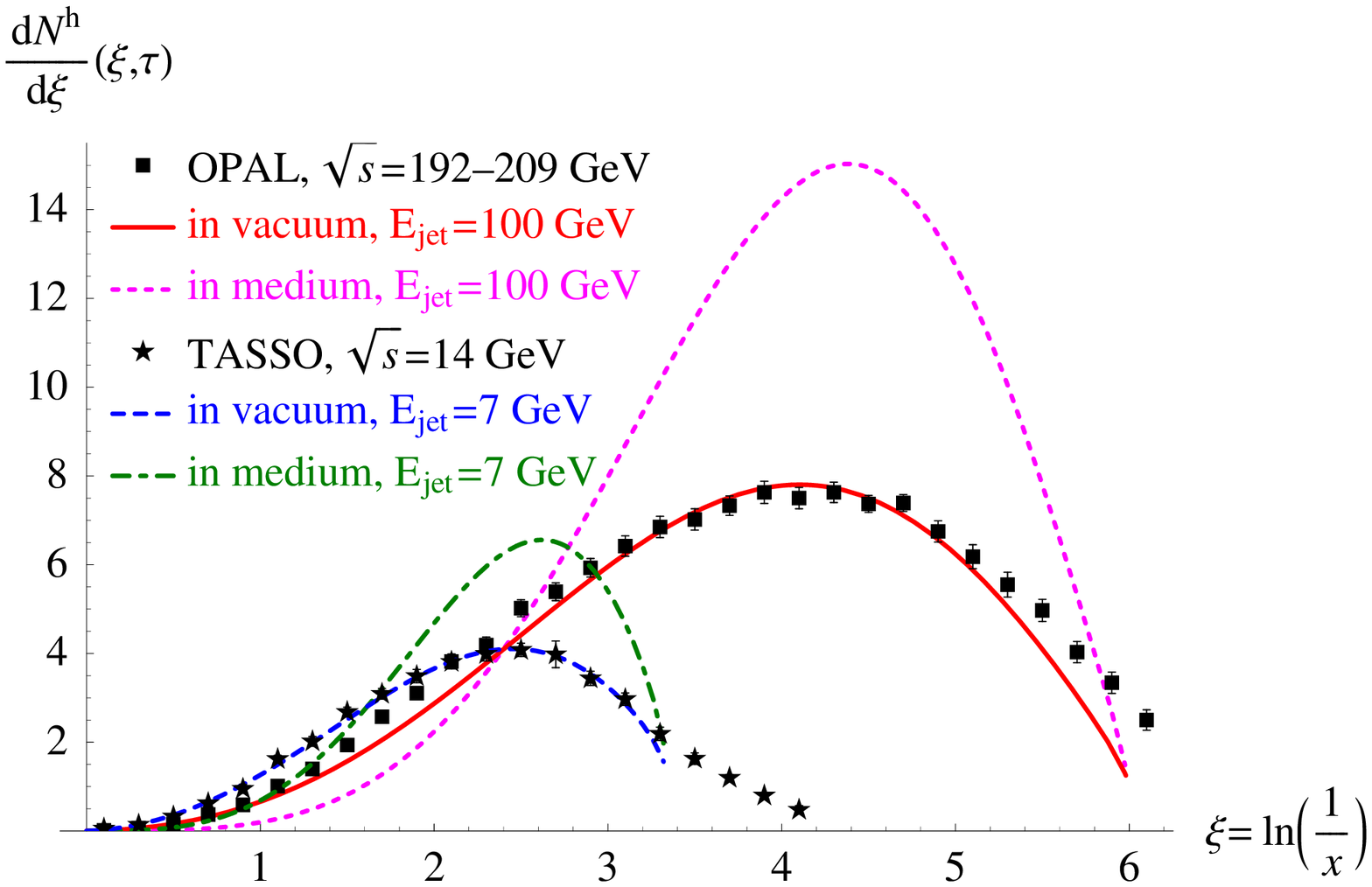}
\caption{\label{Fig:Humpback} MLLA calculations of single inclusive hadron multiplicities compared to e+e- data and for medium modified jets\cite{MLLA}.}
\end{minipage}
\hspace{\fill}
\begin{minipage}[t]{77mm}
\includegraphics[width=\textwidth]{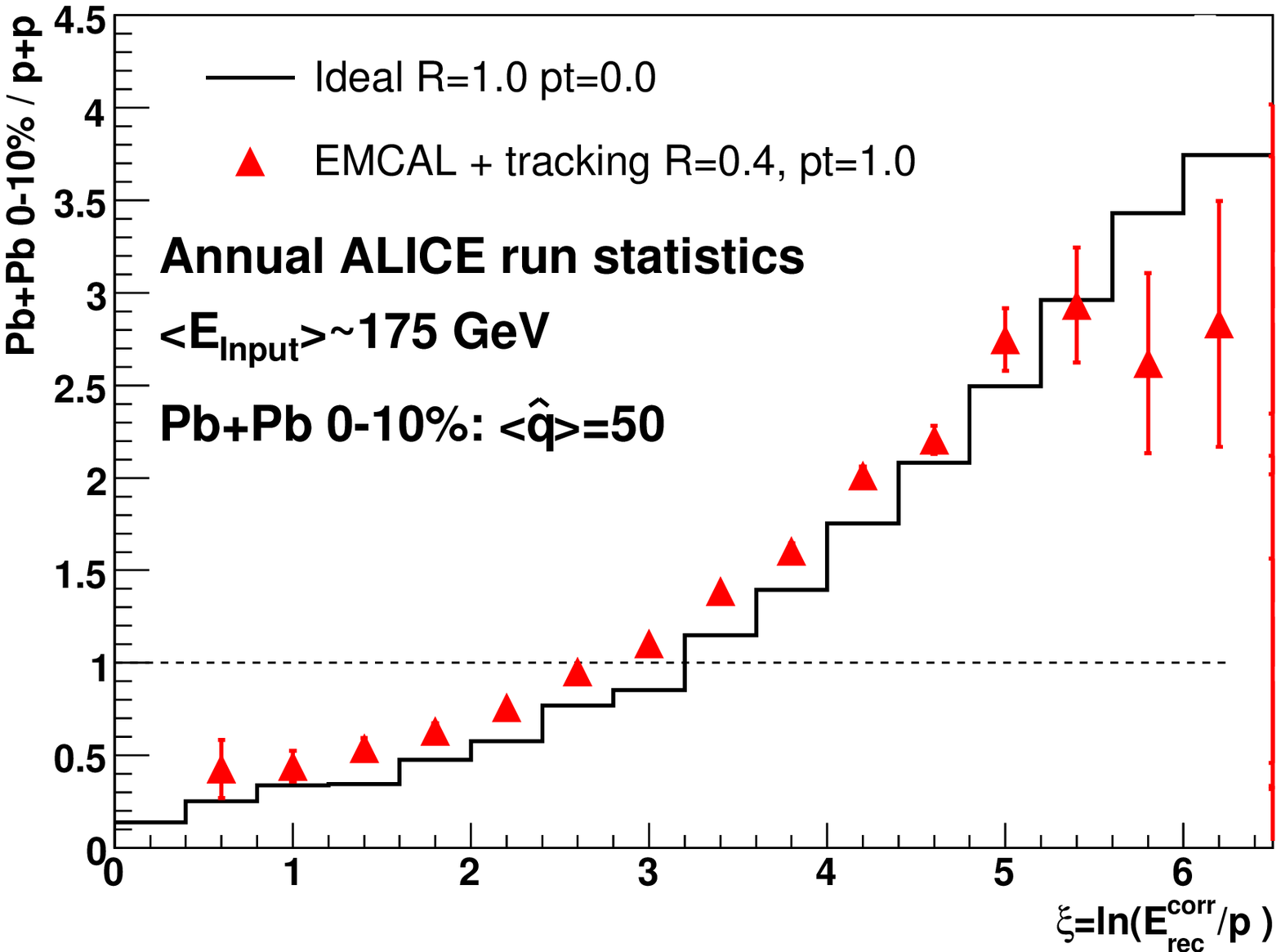}
\caption{\label{Fig:FFEmcal} ALICE expectation for the ratio of quenched $E_T$= 175 GeV jet fragments relative to p+p collisions.  Error bars are $\sqrt{S+B} + 0.002B$.}
\end{minipage}
\end{figure}

Fig.~\ref{Fig:Humpback} shows a recent calculation of the effect of medium-induced gluon radiation on hadron production in the modified leading logarithm approximation (MLLA) parton shower framework\cite{MLLA}.  The MLLA formalism accurately describes the ``hump-backed'' hadron multiplicity distribution vs. $\xi = \ln(E_{jet}/p_{hadron})$ from jet fragmentation in vacuum, as measured in e+e- collisions.  In the presence of a dense medium, the distribution is modified such that low $\xi$ (high $p_T$) hadrons are suppressed while high $\xi$ (low $p_T$) are strongly enhanced.  Even for jets with $E_{T}\sim$ 100 GeV and larger, the excess persists up to $p_{T}\sim$ 5 GeV/c ($\xi\sim$3), which should be measurable above the background even in the highest multiplicity events.

Fig.~\ref{Fig:FFEmcal} shows the expected performance of ALICE for measuring the distortion of the hump-backed plateau for jets with $E_T$=175 GeV in quenched Pb+Pb collisions compared to p+p.
The quenching model used for Pb+Pb simulates the effect of a medium transport coefficient of $\langle\hat{q}\rangle = 50 GeV^{2}/fm$.  The histogram is the idealized and artificial case of a strong quenching without background while the symbols show the expected signal in Pb+Pb using the EMCal + tracking with a cone radius of R=0.4 and $p_T>$1GeV/c.  The deviation of the data from a horizontal line at unity and the size of the error bars indicate the precision within which jet quenching at $E_{jet}$= 175 GeV can be measured by ALICE.  

\section{Conclusion}
ALICE plus the EMCal is well-suited for measuring the full spectrum of jet quenching observables at mid-rapidity expected in heavy ion collisions at the LHC.  The EMCal provides fast triggering that will extend the kinematic reach in jet $E_T$, substantially increasing the dynamic range over which partonic energy loss will be investigated.  Moreover, the additional information provided by the EMCal will substantially reduce biases compared to leading charged particle or charged jet reconstruction and significantly improve the jet energy resolution in ALICE.  With the high precision tracking and particle identification provided by the ALICE tracking system, detailed studies of modified jet fragmentation from 100 MeV to 100 GeV will be possible, making ALICE very well-suited for studying jet quenching in heavy ion collisions.

%%%%%%%%%%%%%%%%%%%%%%%%%%%%%%%%%%%%%%%%%


\begin{thebibliography}{9}

\bibitem{dAuPapers} J. Adams {\it et al.} [STAR Collaboration], Phys. Rev. Lett. 91, 072304 (2003); S.S. Adler {\it et al.} [PHENIX Collaboration], Phys. Rev. Lett. 91, 072303 (2003); B.B. Back {\it et al.} [PHOBOS Collaboration], Phys. Rev. Lett. 91, 072302 (2003); I. Arsene {\it et al.} [BRAHMS Collaboration], Phys. Rev. Lett. 91, 072305 (2003).

\bibitem{BDMPS} R. Baier, Yu. L. Dokshitzer, S. Peigne and D. Schiff, Phys. Lett. B 345, 277 (1995).

\bibitem{GLV} M. Gyulassy, P. Levai and I. Vitev, Phys. Rev. Lett. 85, 5535 (2000).

\bibitem{WangGuo} X.-N. Wang and X.-F. Guo, Nucl. Phys. A 696, 788 (2001).

\bibitem{Zakharov} B. G. Zakharov, JETP Lett. 63 (1996) 952.

\bibitem{SalgadoWiedemann} C. Salgado and U. Wiedemann, Phys. Rev. D 68, 014008 (2003).

\bibitem{STARFuqiang} J. Adams {\it et al.} [STAR Collaboration], Phys. Rev. Lett. 95, 152301 (2005).

\bibitem{STARMagestro} J. Adams {\it et al.} [STAR Collaboration], submitted to Phys. Rev. Lett., nucl-ex/0604018.

\bibitem{MachCone} J. Casalderrey-Solana, E. Shuryak and D. Teaney, Nucl. Phys. A 774, 577 (2006).

\bibitem{Pythia} T. Sjostrand {\it et al.}, Computer Physics Commun. 135, 238 (2001).

\bibitem{Hijing} X. N. Wang and M. Gyulassy, Phys. Rev. D 44, 3501 (1991).

\bibitem{CERNYellowReportJets} A. Accardi {\it et al.}, CERN Yellow Report, hep-ph/0310274.

\bibitem{YellowReportHeavyQuarks} M. Bedjidjian {\it et al.}, CERN Yellow Report, hep-ph/0311048.

\bibitem{CDFbjets} D. Acosta {\it et al.} [CDF Collaboration], Phys. Rev. D 69, 072004 (2004).

\bibitem{KlaySQM04} N. Armesto {\it et al.}, hep-ph/0501225; J.L. Klay, J. Phys. G 31, S1115 (2005).

\bibitem{MLLA} N. Borghini and U.A. Wiedemann, hep-ph/0506218.

\bibitem{SalgadoWiedemannJetShape}  C. A. Salgado and U. Wiedemann, Phys. Rev. Lett. 93, 042301 (2004).

\bibitem{ALICEPPR} F. Carminati {\it et al.} [ALICE Collaboration], J. Phys. G 30, 1517 (2004).

\bibitem{EMCALTP} ALICE Collaboration, CERN-LHCC-2006-014, CERN/LHCC 96-32-Add3 (2006).

\bibitem{UA1Jets} G. Arnison {\it et al.} [UA1 Collaboration], Phys. Lett. B 132, 214 (1983).

\bibitem{JPhysGJets} S.-L. Blyth {\it et al.}, submitted to J. Phys. G, nucl-ex/0609023.

\bibitem{CDFChargedJets} T. Affolder {\it et al.} [CDF Collaboration], Phys. Rev. D 65, 092002 (2002).

\end{thebibliography}
\end{document}